# Magnetic Transition in Monolayer VSe$_2$ *via* Interface Hybridization

Wen Zhang,[1,*] Lei Zhang,[1] Ping Kwan Johnny Wong,[2,*] Jiaren Yuan,[1] Giovanni Vinai,[3] Piero Torelli,[3] Gerrit van der Laan,[4] Yuan Ping Feng,[1] Andrew Thye Shen Wee[1,*]

[1]Department of Physics, National University of Singapore, 2 Science Drive 3, Singapore 117542, Singapore

[2]Centre for Advanced 2D Materials and Graphene Research Centre, National University of Singapore, 6 Science Drive 2, Singapore 117546, Singapore

[3]Laboratorio TASC, IOM-CNR, S.S. 14 km 163.5, Basovizza, 34149 Trieste, Italy

[4]Magnetic Spectroscopy Group, Diamond Light Source, Didcot OX11 0DE, UK

---

[*] E-mails: xiaotur@gmail.com, pingkwanj.wong@gmail.com, phyweets@nus.edu.sg

TOC

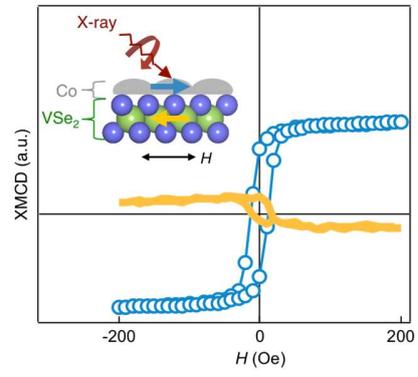

ABSTRACT. Magnetism in monolayer (ML) $VSe_2$ has attracted broad interest in spintronics while existing reports have not reached consensus. Using element-specific X-ray magnetic circular dichroism, a magnetic transition in ML $VSe_2$ has been demonstrated at the contamination-free interface between Co and $VSe_2$. *Via* interfacial hybridization with Co atomic overlayer, a magnetic moment of about 0.4 $\mu_B$ per V atom in ML $VSe_2$ is revealed, approaching values predicted by previous theoretical calculations. Promotion of the ferromagnetism in ML $VSe_2$ is accompanied by its antiferromagnetic coupling to Co and a reduction in the spin moment of Co. In comparison to the absence of this interface-induced ferromagnetism at the Fe/ML-$MoSe_2$ interface, these findings at the Co/ML-$VSe_2$ interface provide clear proof that the ML $VSe_2$, initially with magnetic disorder, is on the verge of magnetic transition.

The discovery of graphene has led to an enormous scientific frenzy in the study of two-dimensional (2D) materials that might show new physical properties ultimately leading to potential new applications. Particularly attractive materials are the transition-metal dichalcogenides (TMDs), due to their naturally layered van der Waals structures and their wide variety of materials properties.[1, 2] Controlling the material thickness down to the single-atom scale provides a powerful tuning of the interacting electronic states and phases, and it is thus highly desirable to fabricate high-quality monolayer (ML) TMDs and to investigate their electronic and magnetic properties.

Recent theoretical predictions have triggered interest in hybrid interfaces between ML TMDs and $3d$ ferromagnets,[3, 4] which is attractive in pushing magnetic devices to the atomically thin limit. However, an experimental demonstration of contamination-free and large-scale ferromagnet/ML-TMD interface has been lacking till recently.[5] In our very recent work, we have demonstrated a growth strategy to create a clean ferromagnet/TMD interface, which enables wafer-scale TMDs on the substrate and the subsequent ferromagnet deposition free from interface contamination.[5] These ensure the interfacial effect under investigation (and particularly the interface-induced magnetism in the monolayers, if any) will not be disturbed by extrinsic impurities.

Here, we present a magnetic phase transition in ML VSe$_2$ *via* interface hybridization with Co, by taking advantage of the abovementioned growth strategy. Using element-specific X-ray magnetic circular dichroism (XMCD), we reveal that such a transition leads to a reliable long-range magnetic order in ML VSe$_2$, with a (total) magnetic moment of about 0.4 $\mu_B$ per V atom, approaching values predicted by previous theoretical calculations. The antiferromagnetic coupling between Co and VSe$_2$, reduced spin moment of Co, and the absence of the interface-induced magnetism in *semiconducting* ML MoSe$_2$ are further proof of this interfacial hybridization effect. It provides evidence that the *metallic* ML VSe$_2$, initially with magnetic disorder, is on the verge of magnetic transition.

RESULTS AND DISCUSSION

**Structural characterization and XA/XMCD spectroscopies of ML VSe$_2$.** **Figure 1**a shows a typical scanning tunneling microscopy (STM) image of ML VSe$_2$ grown on HOPG. The individual VSe$_2$ terraces are clearly identified by well-defined hexagonal or triangular edges, with the average size of the order of 100 nm. The line profile drawn across the terrace indicates a step height of ~7 Å, in agreement with prior reported values.

To verify the predominantly $d^1$ configuration in our ML VSe$_2$, we perform X-ray absorption (XA) spectroscopy at the V $L_{2,3}$ edge, with the incidence angle of the X-ray beam set to 45° relative to the sample surface normal (**Figure 1**b). As shown in the upper panel of **Figure 1**c, two main peaks are revealed at ~516 and ~523 eV, corresponding to the V $L_3$ and $L_2$ absorption edges due to dipole-allowed transitions from the spin-orbit split V $2p_{3/2}$ and $2p_{1/2}$ core levels to the $3d$ unoccupied states. The shoulders at energies below those of the main peaks, as indicated by the arrow, are remnant of the atomic multiplet structure of the $d^1$ configuration. These spectral features are akin to those measured from other $3d^1$ vanadium compounds,[6, 7] thus providing a spectroscopic fingerprint of the $1T$ phase of the monolayer.

To characterize the *intrinsic* magnetic state in the VSe$_2$, we have performed XMCD measurement at the V $L_{2,3}$ edge, by recording at remanence after reversing the direction of the external magnetic field (±500 Oe) applied in the sample plane at each energy point. As shown in the lower panel of **Figure 1**c, the XMCD contrast, within the scale of experimental error, is negligible and thus rules out the existence of intrinsic ferromagnetism in the monolayer. This is consistent with our previous work, demonstrating ML VSe$_2$ as a frustrated magnet, in which its spins exhibit subtle correlations albeit in the absence of a long-range magnetic order.[8]

**Long-range ferromagnetic order in ML VSe$_2$ induced by direct contact with cobalt.** After removal of the Se cap by *in-situ* thermal annealing, Co overlayers are deposited on the clean surface of the ML VSe$_2$, as shown in **Figure 2**a. **Figure 2**b shows the corresponding STM image, revealing a 3D growth of Co. The V $L_{2,3}$ XMCD spectra in the lower panel of **Figure 2**c highlight the emergence of considerable magnetic polarization on the V atoms upon 3-ML Co deposition. The sign of the V $L_{2,3}$ is opposite to that of the Co $L_{2,3}$ XMCD in the upper panel, indicating an antiparallel alignment between the spins in ML VSe$_2$ and Co. Note that both the Co and V XMCD signals are enhanced when cooling down to about 65 K. The antiparallel alignment between the Co and V magnetic moments suggests the antiferromagnetic coupling between Co and VSe$_2$, further confirmed by element-specific XMCD hysteresis loops, as displayed in **Figure 2**d. The high squareness and remanence ratio of the V loop are clear proofs of its ferromagnetic character.

From the XMCD spectra, the spin- ($m_S$) and orbital- ($m_L$) magnetic moments of Co and V are extracted using the sum-rule analysis (see Supplementary Materials **Note 1** and **Figure S1**).[9, 10] As shown in Table 1, the orbital moment of V, $m_{L,V}$, carries the opposite sign as the spin moment, $m_{S,V}$, in agreement with the Hund's rule for a less than half-filled $3d$ shell. Moreover, the resulting total moment of V, $m_{tot,V} = m_{L,V} + m_{S,V}$, has an opposite sign to that of Co ($m_{tot,Co}$), confirming an antiferromagnetic coupling between Co and ML VSe$_2$, as mentioned above.

Interestingly, in 3-ML Co capped VSe$_2$, the $m_{tot,V}$ was lifted to ~0.3 $\mu_B$ at room temperature and ~0.4 $\mu_B$ at 65 K, approaching the theoretical values at absolute zero.[11, 12] This demonstrates a transition from disorder to long-range ferromagnetic order in ML VSe$_2$ at the interface. A similar transition from paramagnetism to ferromagnetism has been reported at the Co/V interface,[13] which was interpreted as a direct consequence of Co-V hybridization across the interface. Similarly, in the present case of Co/ML-VSe$_2$, an interfacial hybridization induced transfer of the exchange splitting between the two materials leads to a splitting of the spin subbands in the V with the spin-down band having a lower energy, as shown in the calculated results in the upper panel of **Figure 3**a. Such induced exchange splitting results in an induced spin moment that is coupled antiparallel to the Co moment, as reproduced by theoretical calculations in the lower panel of **Figure 3**a, where the exchange splitting of V is opposite to that of Co at the interface. Such an antiferromagnetic coupling is found to be energetically favorable comparing to the case of parallel alignment at the interface, in particular between the more-than-half-filled (like Co 3$d$) and less-than-half-filled (like V 3$d$) bands.[14] We have also observed that the magnetic transition only arises when the Co overlayer becomes ferromagnetic (see **Figure S2**), further evidencing the major role played by the abovementioned interfacial magnetic hybridization compared with the chemical transition (discussed below) in ML VSe$_2$ induced by the Co deposition. It is noteworthy that the $m_{S,Co}$ at the interface is smaller than that of bulk Co,[15] probably caused by the exchange-splitting transfer from Co to ML VSe$_2$. Such a transfer may push down the Co spin-down band, and in turn reduce the magnitude of its exchange splitting, in line with the theoretical calculation shown in the lower panel of **Figure 3**a. A similar scenario has also been observed at the Co/V interface as well as for the Fe magnetic moment in the V/Fe system,[13, 16, 17] thus serving as another proof of the hybridization effect.

The $m_{L,V}/m_{S,V}$ ratio is slightly below 2%, which is consistent with the previous calculation for VFe alloy.[17] For the Co/ML-VSe$_2$ interface, the $m_{L,Co}$ and the $m_{L,Co}/m_{S,Co}$ ratio are found to be slightly higher than those at the Co/graphite interface.[18] This might be caused by the stronger hybridization-induced transfer of the spin-orbit splitting between Co and VSe$_2$ than between Co and graphite, where the latter substrate features a much weaker spin-orbit interaction. The orbital moment depends strongly on the orbital symmetry of the spin-split states near the Fermi level within a typical energy range determined by the effective $d$ spin-orbit interaction.[19].

**Lacking of interface-induced ferromagnetism at the Fe/ML-MoSe$_2$ interface.** For comparison, we have prepared a similar hybrid interface between Fe and ML MoSe$_2$ (**Figure 4**a). Here Fe replaces Co as the ferromagnetic layer to avoid any spurious contribution on the Mo $M_{2,3}$-edge XMCD spectra from the $L_{2,3}$ edges of Co. Specifically, the Mo $M_{2,3}$ edges lie at the second order of the plane grating monochromator of Co $L_{2,3}$ edges. This consequently leads to a non-trivial data treatment for both XAS and XMCD of Mo. Therefore, Fe was chosen instead, since its half energy does not overlap with the Mo edges. The STM image in **Figure 4**b shows the 3D growth of Fe on the ML MoSe$_2$, similar to the case of Co on ML VSe$_2$. Upon deposition of Fe, we see both the Mo $3d$ and Se $3d$ XPS shift to a lower binding energy, accompanied with the appearance of a small shoulder (marked by the arrows in **Figure 4**c). This shoulder suggests new bonding states probably derived from Mo-Fe at the interface.[20] With increasing coverage of Fe, the shifts were enhanced and the shoulder becomes more apparent. These were also accompanied by the decreasing XA intensity at the Mo $M_{2,3}$ edges (**Figure S3**). Strikingly, however, the interface-induced magnetism was not observed at the Fe/ML-MoSe$_2$ interface. As shown in **Figure 4**d, the 8-ML Fe has shown strong ferromagnetism with a total moment of about 1.98 $\mu_B$ at 55 K, while the Mo XMCD still remains absent. The lack of interface-induced magnetism in ML-MoSe$_2$ could be attributed to a lack of DOS at the Fermi level. The contrasting observation at the Co/ML-VSe$_2$ and Fe/ML-MoSe$_2$ interfaces again serves as strong evidence for the interfacial hybridization effect demonstrated above, which is robust in metallic TMDs (such as ML VSe$_2$) while practically negligible in semiconducting TMD (*e.g.*, ML MoSe$_2$).

**Charge transfer at the Co/ML-VSe$_2$ interface.** Apart from the abovementioned exchange-splitting transfer from Co to ML VSe$_2$, a charge transfer at the interface should also be considered. As shown in **Figure 5**a, upon deposition of the first ML Co, the V $L_{2,3}$ XA peaks shift to lower photon energies, suggesting that the V ions become less positive. Simultaneously, the XA peaks change from the typical localized V$^{4+}$ multiplet structure to a more metallic shape with smooth asymmetric shape tailing off at the high photon energy side.[21] This change in electronegativity is confirmed by XPS. As shown in **Figure 5**b and **5**c, the pristine ML VSe$_2$ shows a more localized structure. Upon Co deposition, the V $2p$ XPS peak shape becomes more metallic, *i.e.*, shifting to lower binding energy and typical smooth asymmetric shape tailing off at the high binding-energy side. These spectral changes are concomitant with the Se $3d$ peaks shifting to higher binding energies, along with the merging of the initially separated $3d^{5/2}$ and $3d^{3/2}$ peaks. This might suggest that the Se ions of the monolayer accept electrons from the Co overlayer and results in the formation of Co-Se at the interface,[22] as also suggested by theoretical calculations in **Figure 3**b. Note that the XA and XPS energy shifts and spectral changes mainly appear upon the first ML Co, while only slightly changed for the subsequent increase of the Co coverage. This indicates that the charge transfer, if occurred, is confined to the neighboring Co and Se at the interface. It is also worthwhile that no multiplet structure could be revealed in the Co XA and XPS spectra, thus supporting the metallic nature of Co layer at the Co/ML-VSe$_2$ interface even at the single layer limit, as shown in **Figure S4**.

CONCLUSIONS

A recent experimental study indicated that a ferromagnetic state persists up to room temperature in ML VSe$_2$,[23] which, however, raised several critical unexplained features. The most debatable finding is the reported magnetization as large as 15 $\mu_B$ per V atom, which is in strong disagreement with the DFT calculations of bulk and ML VSe$_2$.[12, 24] On the other hand, Feng *et al.* showed by using XMCD that there is a zero magnetic moment on the V atoms down to 10 K in an applied magnetic field of 9 T.[25] Several studies have followed by aiming at solving this ongoing controversy. For instance, Fumega and Pardo reported that DFT calculations yield a ferromagnetic ground state,[26] which is on the verge of instability. They demonstrated that the structural rearrangement due to the charge density wave state causes an energy gap opening at the Fermi level and the quenching of ferromagnetism in the ML limit, offering a possible clue in explaining the above controversy. Our findings are positioned to provide an additional clue in tackling this issue, with an unambiguous demonstration of magnetic transition from disorder to long-range order in our ML VSe$_2$. From the application perspective, the magnetic interface of Co/ML-VSe$_2$ could be intriguing for many spin-based effects, such as spin pumping and spin injection, in magnetic sensor and data storage technologies.

METHODS

**Sample preparation.** ML VSe$_2$ and MoSe$_2$ samples were grown in an in-house-built molecular-beam epitaxy (MBE) chamber, with base pressure of better than $1 \times 10^{-9}$ mbar. HOPG substrates were prepared by *in-situ* cleavage and annealing at 550 °C for at least 5 hours. High-purity V, Mo, and Se were evaporated from separate electron-beam evaporators and a standard Knudsen cell, respectively. The flux ratios of V:Se and Mo:Se were both controlled to be >1:10. During the growth process the substrate temperature was kept at ~400°C. To protect the sample from environmental contamination during *ex-situ* transport through air to other ultrahigh-vacuum (UHV) measurement chambers, a Se capping layer with a thickness of ~20 nm was deposited onto the sample surface after growth. For subsequent characterization by XPS and XA/XMCD, the Se layer was desorbed in UHV at 200°C and transferred to the analysis chamber, where the XPS/XA V 2*p*, Mo 3*d*, and Se 3*d* core-level spectra were immediately measured. Then the sample was transferred back to the preparation chamber for deposition of the Co overlayer. The above procedure, *i.e.*, overlayer deposition of Co and XPS/XA core-level measurements, was repeated until the deposited Co becomes ferromagnetic at room temperature and the underneath VSe$_2$ shows XMCD signal at the V edge. Note that no Se cap was covered on top of Co during the *in-situ* growth and measurement.

**STM/STS characterizations.** STM measurements were carried out in a multichamber ultrahigh-vacuum system housing an Omicron LT-STM interfaced to a Nanonis controller. The base pressure was better than $10^{-10}$ mbar. A chemically etched tungsten tip was used, and the sample was kept at 77 K during the measurement. STM images were recorded in constant-current mode. For the *dI/dV* spectroscopic measurements, the tunneling current was obtained using the lock-in technique with a bias modulation of 40 mV at a frequency of 625 Hz.

**Synchrotron-radiation measurements.** XPS, XA and XMCD measurements were carried out at the High-Energy branch of Advanced Photoelectric Effect experiments (APE-HE) at Elettra Sincrotrone Trieste, Italy, which covers the photon energy range from 150 to 1600 eV.[27] For the XPS at room temperature, an Omicron EA-125 spectrometer was used to collect the spectroscopic data at an incident angle of 45° of the X-ray beam relative to the sample normal. The binding energy of the data was calibrated using the 4*f* core levels and Fermi edge of a reference Au foil. The XA/XMCD spectra were collected in total-electron-yield mode, with the sample drain current recorded as a function of the photon energy, in the presence of an external magnetic field. XMCD were recorded at room temperature and 65 K, using a fixed helicity of 75% circularly polarized X-rays, altering magnetic field pulses of ±500 Oe along the surface plane at each energy point of the spectra. Element-specific hysteresis loops were measured by switching the photon helicity and recording the peak heights of the Co and V $L_{2,3}$ edges and pre-edges, respectively, as a function of the applied magnetic field.

**DFT calculation.** The electronic structure calculations are performed by using the Vienna *ab initio* Simulation Package, (VASP), which is within the framework of DFT.[28] The spin-polarized DFT calculations are done by employing a plane-wave basis set with a cutoff energy of 450 eV.[29] Perdew-Burke-Ernzerhof (PBE) within general gradient approximation (GGA) is chosen as exchange correlation potential.[30] The Co/VSe$_2$ heterostructure is simulated by hexagonal sheet geometry with three atomic layers of Co placed on VSe$_2$ unit cell. A $15 \times 15 \times 1$ $k$-point mesh is utilized and the convergence criterion of the total energy is set to be $10^{-5}$ eV. The structures are fully relaxed until the force is less than 0.01 eV/Å for all atoms.

ACKNOWLEDGMENTS

We acknowledge financial support from the Singapore Ministry of Education Tier 2 grants MOE2016-T2-2-110 and A*STAR Pharos R-144-000-359-305. This work was partially performed in the framework of the Nanoscience Foundry and Fine Analysis (NFFA-MIUR Italy Progetti Internazionali) project.

FIGURES

**Figure 1.** STM topography and XA/XMCD spectroscopies of ML VSe$_2$. (a) STM topography of a typical region of ML VSe$_2$ on a HOPG substrate. The scale bar is 50 nm. The line profile drawn across the terrace indicates a step height of ~7 Å. (b) Crystal structure of 1$T$-VSe$_2$ (side view) and schematic diagram of the experimental geometry for the XA/XMCD measurements. (c) XA spectra (upper) of the ML VSe$_2$, taken at the V $L_{2,3}$ edges, with opposite applied magnetic fields of ±500 Oe. Their difference (μ$^-$-μ$^+$) gives the XMCD spectrum (lower).

**Figure 2.** Long-range ferromagnetic order in ML VSe$_2$ induced by direct contact with cobalt. (a) Schematic illustration of the antiferromagnetic coupling between Co and VSe$_2$. (b) STM topography of ~5 ML Co deposited on ML VSe$_2$. The VSe$_2$ surface is fully covered by Co, with dome-shaped features arising from the 3D growth of Co extending to the whole terrace surface of the VSe$_2$. (c) XMCD spectra taken at Co (upper) and V (lower) $L$ edges, revealing an induced ferromagnetism in ML VSe$_2$ at room temperature arising from the proximity of a 3-ML Co. The XMCD signal (lower panel) highlights the considerable magnetic polarization on the V atoms, with a sign opposite to that of the Co, indicating that the induced moments in ML VSe$_2$ are coupled antiparallel to Co in the film plane. (d) XMCD hysteresis loops obtained at Co and V $L_3$ edge at 65 K, which confirm the antiparallel alignment, *i.e.* antiferromagnetic coupling, between Co and V spins.

**Figure 3.** (a) Projected density of states (PDOS) of V (upper, red line, given the primary contribution of the V 3$d$ orbitals to the DOS around the Fermi level of ML VSe$_2$.[12]) and Co (lower, dark blue line) at the Co/ML-VSe$_2$ interface. The calculation consistently reproduces our experimental observation on the antiferromagnetic coupling between VSe$_2$ and Co. The PDOS of 3-ML Co before deposition onto ML VSe$_2$ is also shown in the lower panel (in light blue color) for comparison, and we see that the inclusion of ML VSe$_2$ leads to a shift of the Co spin-down band to lower energy and thus a reduction in the exchange splitting. This in turn results in a reduction in the Co spin moment. (b) The differential charge densities were calculated from the difference between the charge density of Co/VSe$_2$ heterojunction and the summed charge density of the ML VSe$_2$ and Co layer. The gray, purple, and green spheres represent Co, Se, and V atoms, respectively. The charge accumulation (yellow isosurface) near the Se atom and the charge depletion (aqua isosurface) near Co atom suggest a charge transfer from Co layer to ML-VSe$_2$. Also the electron accumulation between the Co and Se atoms at the interface implies the formation of the Co-Se bond.

**Figure 4.** XMCD for the Fe/ML-MoSe$_2$ interface. (a) Lattice structure of 2$H$-MoSe$_2$ (side view). (b) STM topography of Fe (bright spots) deposited on ML MoSe$_2$ (terrace). (c) XPS of Mo and Se upon deposition of Fe. The arrows mark the shoulder feature arising from Fe-Mo bonding states. (d) XA/XMCD spectra taken at the Fe $L_{2,3}$ and Mo $M_{2,3}$ edges at 55 K. No magnetic signal is observed in MoSe$_2$ with 8-ML Fe coverage.

**Figure 5.** XA and XPS spectra on the Co/ML-VSe$_2$ interface. (a) Summed XA spectra taken at the V $L_{2,3}$ edge. Upon deposition of the first ML Co, the V $L_{2,3}$ XA peaks shift to lower photon energies, suggesting that the V ions become less positive, accompanied by disappearance of the localized V$^{4+}$ multiplet structure. (b) XPS of V 2$p$ and Se 3$d$. Upon deposition of the first ML Co, the V 2$p$ XPS peaks shift to lower binding energies, concomitant with the shift of Se 3$d$ peaks to higher binding energies and the merging of the initially existing double peaks. Note that the XA and XPS energy shifts and spectral changes mainly appear upon the first ML Co, while only slightly changed for the subsequent increase of the Co coverage, thus indicating that the charge transfer, if occurred, is confined to the neighboring Co and Se at the interface.

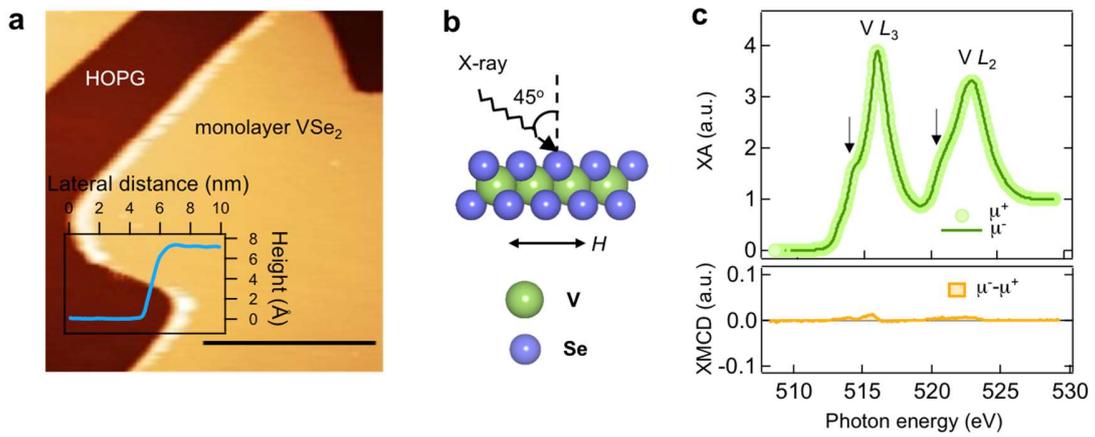

Figure 1

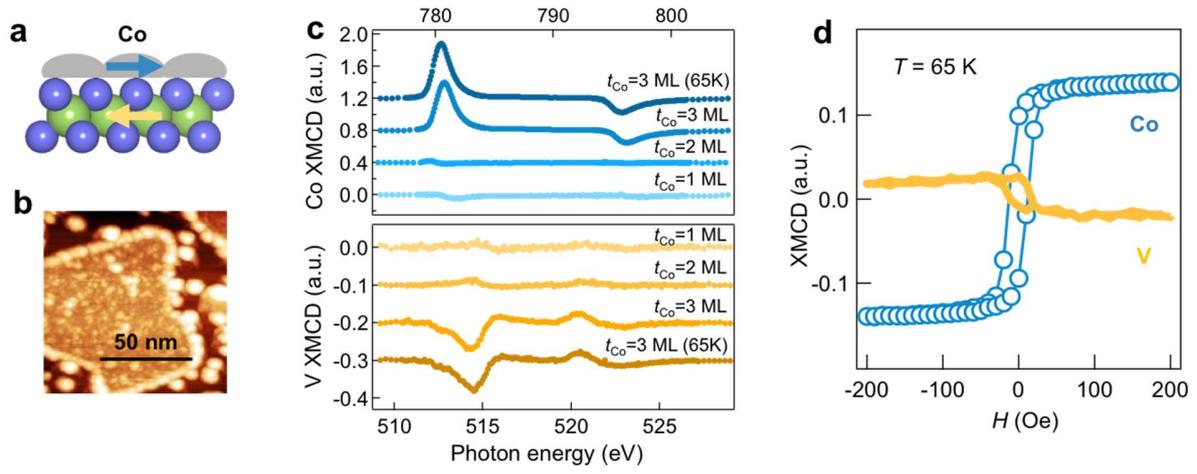

Figure 2

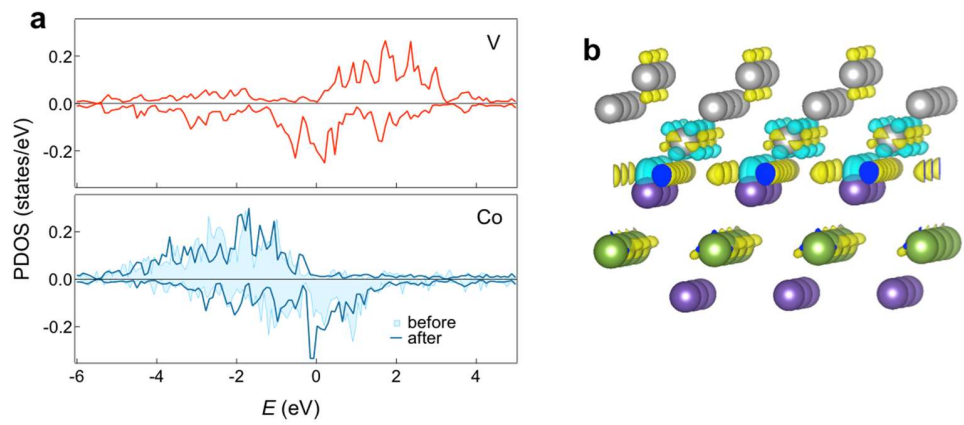

Figure 3

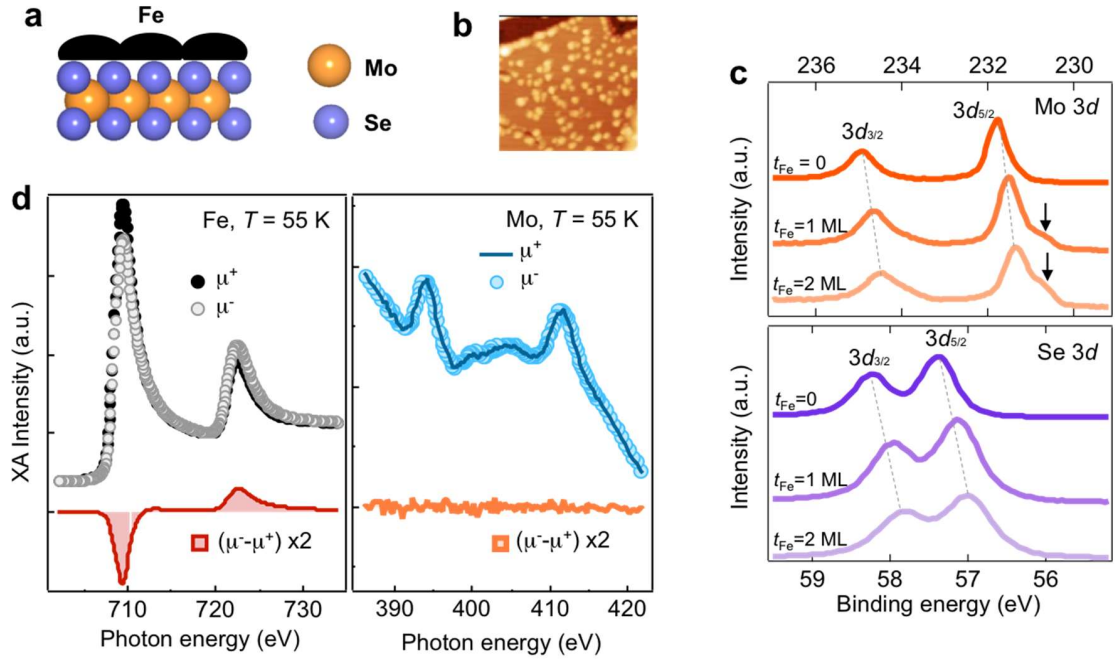

Figure 4

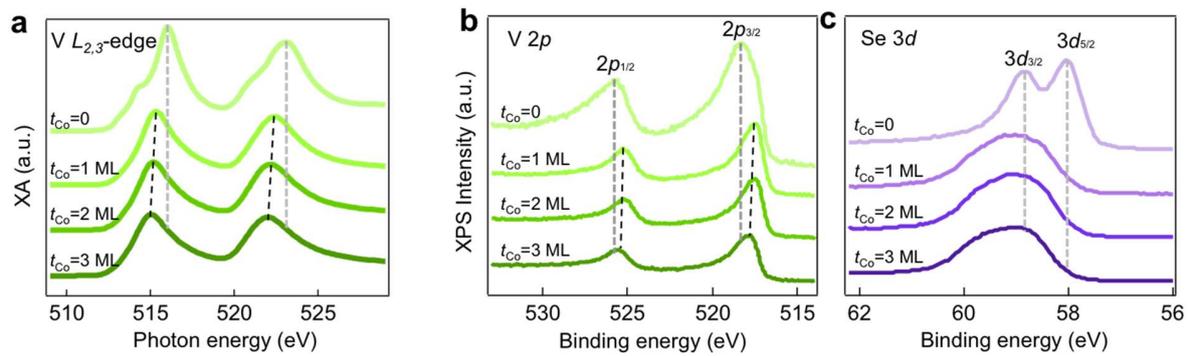

Figure 5

TABLES

**Table 1.** Microscopic magnetic moments at the Co/ML-VSe$_2$ interface. The negative sign of $m_{S,V}$ indicates its antiparallel alignment to $m_{L,V}$, and the negative $m_{tot,V}$ represents its antiparallel alignment to $m_{tot,Co}$. The error bar generated in the sum rules analysis for the Co moment is about 10%, and that for the V spin moment ($m_{S,V}$) and its related values should be larger than 10%.

| | $m_{L,Co}$ ($\mu_B$/hole) | $m_{S,Co}$ ($\mu_B$/hole) | $m_{tot,Co}$ ($\mu_B$/hole) | $m_{L,Co}/m_{S,Co}$ | $m_{L,V}$ ($\mu_B$/atom) | $m_{S,V}$ ($\mu_B$/atom) | $m_{tot,V}$ ($\mu_B$/atom) | $m_{L,V}/m_{S,V}$ |
|---|---|---|---|---|---|---|---|---|
| 3-ML Co/ML-VSe$_2$ | 0.120 | 0.47 | 0.590 | 0.26 | 0.004 | -0.30 | -0.30 | -0.013 |
| 3-ML Co/ML-VSe$_2$ (69 K) | 0.124 | 0.55 | 0.674 | 0.23 | 0.007 | -0.40 | -0.40 | -0.018 |
| 7-ML Co/1.8-ML V[13] | 0.107 | 0.69 | 0.797 | 0.16 | - | - | -1.2 | - |
| Bulk Co[15] | 0.067 | 0.71 | 0.777 | 0.10 | - | - | - | - |
| Co/graphite (77 K)[18] | 0.096 | 0.44 | 0.536 | 0.22 | - | - | - | - |

ASSOCIATED CONTENT

**Supporting Information Available:** XMCD sum rules analysis; XMCD hysteresis loops of Co grown on ML VSe$_2$; Mo $M$-edge XA spectra with increasing Fe coverage in Fe/ML-MoSe$_2$; Co $L$-edge XA and XPS spectra after deposition on ML VSe$_2$. (PDF) This material is available free of charge *via* the Internet at http://pubs.acs.org.

The authors declare no competing financial interests.

AUTHOR INFORMATION

**Corresponding Author**

* E-mail: xiaotur@gmail.com.

* E-mail: pingkwanj.wong@gmail.com.

* E-mail: phyweets@nus.edu.sg.